\documentclass{llncs} 
\usepackage{times}

\usepackage{algpseudocode}
\usepackage{algorithm}
\usepackage{amsmath}
\usepackage{amsfonts}
\usepackage{amssymb}
\usepackage{bbm}
\usepackage{booktabs}
\usepackage{caption}
\usepackage{capt-of}
\usepackage{cite}
\usepackage{color}
\usepackage{float}
\usepackage{graphicx}
\usepackage{mdwlist}    % from christos / mary: makes lists tighter
\usepackage{multirow}
\usepackage{tikz}
\usepackage{wrapfig}
\usepackage{subfigure}
\usepackage{url}
\usepackage{layouts}

\usetikzlibrary{arrows}

\def\pprw{8.5in}
\def\pprh{11in}

\newcommand{\hide}[1]{}

\newcommand{\bit}{\begin{itemize*}}
\newcommand{\eit}{\end{itemize*}}
\newcommand{\ben}{\begin{enumerate*}}
\newcommand{\een}{\end{enumerate*}}

\newcommand{\model}{\textsc{Phoenix-R} }

\newcommand\blfootnote[1]{%
    \begingroup
    \renewcommand\thefootnote{}\footnote{#1}%
    \addtocounter{footnote}{-1}%
    \endgroup
}

\begin{document}
\title{Revisit Behavior in Social Media: \\ The Phoenix-R Model and Discoveries}

\author{Flavio Figueiredo\inst{1} \and Jussara M. Almeida\inst{1} \and \\ Yasuko Matsubara\inst{2,3} \and Bruno Ribeiro\inst{3} \and Christos Faloutsos\inst{3}}

\institute{Department of Computer Science, Universidade Federal de Minas Gerais
\and 
Department of Computer Science, Kumamoto University
\and
Department of Computer Science, Carnegie Mellon University}

%for sizing figures
%347.0pt is the colwidth
%\showthe\columnwidth

\maketitle

\begin{abstract}
How many listens will an artist receive on a online radio? How about plays on a YouTube video? How many of these visits are new or returning users? Modeling and mining popularity dynamics of social activity has important implications for researchers, content creators and providers. We here investigate the effect of revisits (successive visits from a single user) on content popularity. Using four datasets of social activity, with up to tens of millions media objects (e.g., YouTube videos, Twitter hashtags or LastFM artists), we show the effect of revisits in the popularity evolution of such objects. Secondly, we propose the \model model which captures the popularity dynamics of individual objects. \model has the desired properties of being: (1) parsimonious, being based on the minimum description length principle, and achieving lower root mean squared error than state-of-the-art baselines; (2) applicable, the model is effective for predicting future popularity values of objects.
% \svnidlong
% {$HeadURL: svn+ssh://christos@mcfarland.db.cs.cmu.edu/usr4/SVN-papers/christos-papers/samplePaper6svn/000abstract.tex $}
% {$LastChangedDate: 2010-01-17 20:24:41 -0500 (Sun, 17 Jan 2010) $}
% {$LastChangedRevision: 409 $}
% {$LastChangedBy: christos $}
% \svnRegisterAuthor{christos}{Christos Faloutsos}

\end{abstract}

\section{Introduction}
\label{sec:intro}
{\it How do we quantify the popularity of a piece of content in social media applications? Should we consider only the audience (unique visitors) or include revisits as well? Can the revisit activity be explored to create more realistic popularity evolution models? } These are important questions in the study of social media popularity. In this paper, we take the first step towards answering them based on four large  traces  of user activity collected from different social media applications: Twitter, LastFM, and YouTube\footnote{\url{http://twitter.com} \quad \url{http://lastfm.com} \quad \url{http://youtube.com}}.

Understanding the popularity dynamics of online content is both a challenging task,  due to the vast amount and variability of content available, as it can also provide invaluable insights into  the behaviors of human consumption~\cite{Crane2008} and into more effective engineering strategies for online services. A large body of previous work has investigated the popularity dynamics of social media content,  focusing mostly on modeling and predicting the {\it  total number of accesses} a piece of content receives~\cite{Li2013,Figueiredo2011,Cha,Crane2008,Radinsky2012}. 

\begin{figure}[t]
 \centering
 \subfigure[Rock Song (growth in popularity)]{\includegraphics[scale=.7]{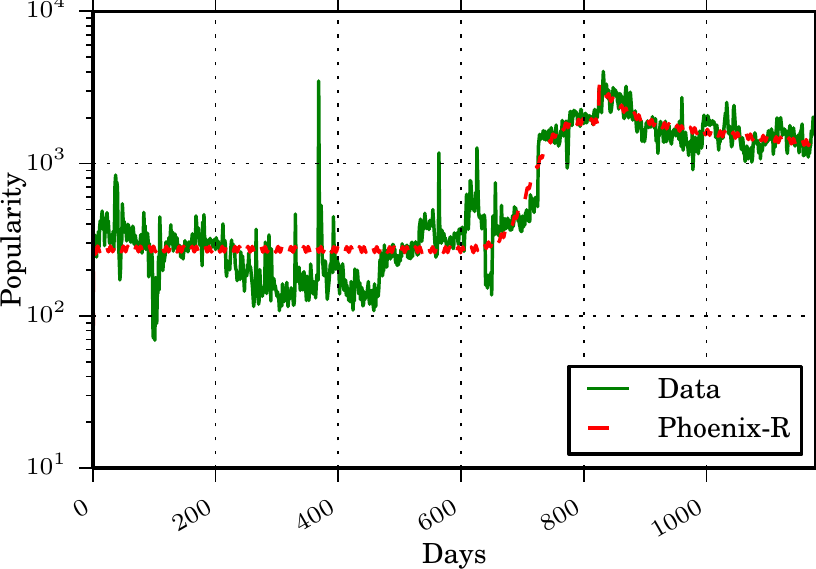}}
 \subfigure[Flashdance (80's movie) clip (revisits)]{\includegraphics[scale=.7]{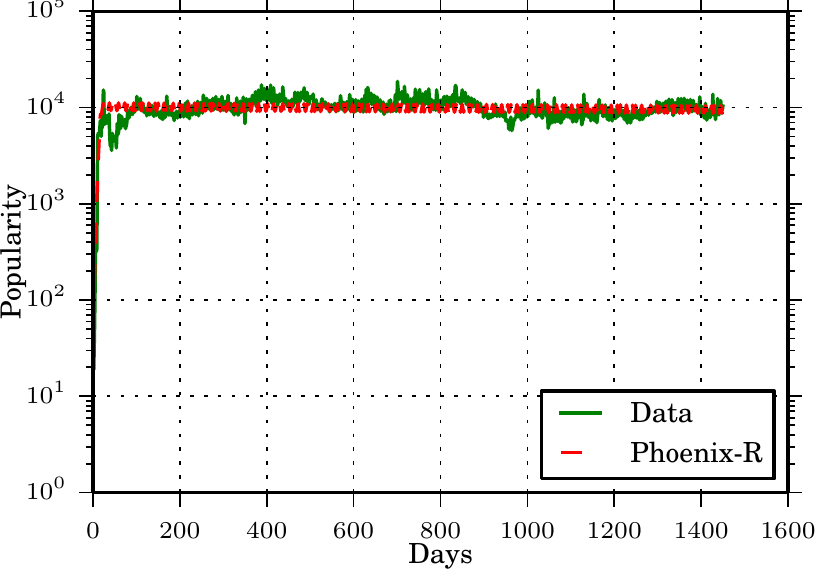}}
 \subfigure[Korean Music Video (single cascade)]{\includegraphics[scale=.7]{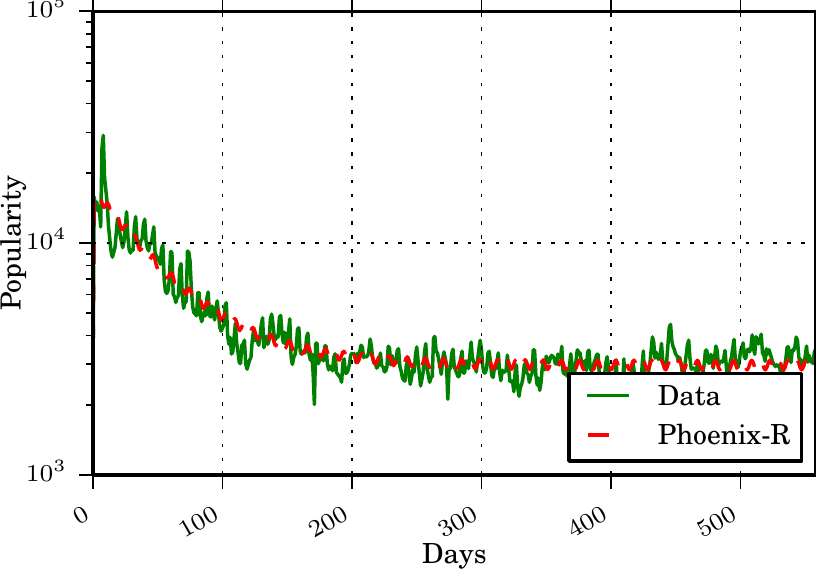}}
 \subfigure[User Dancing Video (single cascade)]{\includegraphics[scale=.7]{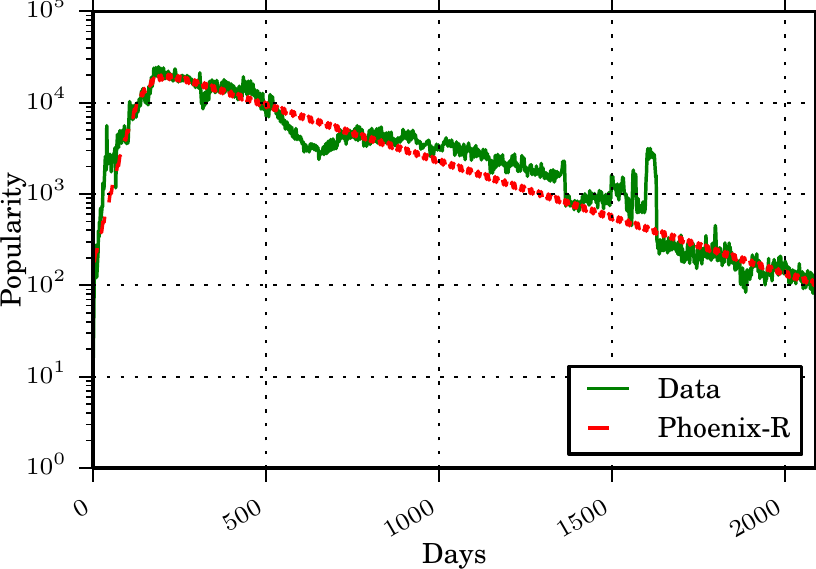}}
 \caption{Different YouTube videos as captured by the \model model.}
 \label{fig:ex}
 \vspace{-.8cm}
\end{figure}

However, a key aspect that has not been explored by most previous work is the effect of revisits on content. The distinction between audience (unique users), revisits (returning users), and popularity (the sum of the previous two) can have large implications for different stakeholders of these applications - from content providers to content producers - as well as for internal and external services that rely on social activity data. For example, marketing services should care most about the audience of a particular content, as opposed to its total popularity, as each access does not necessarily represent a new exposed individual. Even system level services, such as geographical sharding~\cite{Duong2013,Vakali2012}, can be affected by such distinction, as a smaller audience served by one data center does not necessarily imply that a smaller volume of activity (and thus lower load) should be expected.  As prior studies of content popularity in social media  do not  clearly distinguish between unique and returning visits, the literature still lacks fundamental  knowledge about content popularity dynamics in this environment.

{\bf Goals:} We here aim at investigating and modeling the effect of revisits on popularity, thus complementing prior efforts on the field of social media popularity. Our goals are:
(1)  Characterizing the revisits phenomenon and show how it affects the evolution of popularity of different objects (videos, artists or hashtags) on social media applications; (2) Introducing the \model model that captures the evolution of popularity of individual objects, while explicitly accounting for revisits. Also, we develop the model so that it can capture multiple cascades, or outbreaks, of interest in a given object.

{\bf Discoveries: } Among other findings, we show that when analyzing total popularity values, revisits account from 40\% to 96\% of the popularity of an object (on median), depending on the application. Moreover, when looking at small time windows (e.g., hourly) revisits can be up to 14x more common than new users accessing the object.

{\bf \model Results: } The \model model explicitly addresses revisits in social media behavior and is able to automatically identify multiple cascades~\cite{Hu2014} using only popularity time series data. This is in contrast to previous methods such as the  SpikeM~\cite{Matsubara2012} approach, which models single cascades only, and the TemporalDynamics~\cite{Radinsky2012} models, are linear in nature. Figure~\ref{fig:ex} shows the different behaviors which can be captured by the Phoenix-R model. Notice how the model captures a growth in the popularity of video (a), videos which have a plateau like popularity after the upload (b), and two different single cascade dynamics (c-d). The \model model is also scalable. Fitting is done in linear time and no parameters are required.

%In comparison to previous methods, the SpikeM~\cite{Matsubara2012} was able to model single cascades only. Moreover, the TemporalDynamics~\cite{Radinsky2012} models, are linear in nature. \model is defined as sum of multiple independent cascades, or shocks. Each cascade accounts for factors not captured by previous models (e.g., multiple revisits and user loss of interest on content). To achieve a parsimonious description of the data, we fit the model based on minimum description length principle (MDL), achieving over 10x lower root mean squared errors than baseline methods depending on the dataset. 

%models when describing the popularity time series. 

%Showing the accuracy of our model, and that revisits and multiple cascades should be taken into account for long term popularity modeling of social media. 

{\bf Outline:} Section~\ref{sec:background} presents an overview of definitions and background. This is followed by Section~\ref{sec:charact} which presents our characterization. \model is described in Section~\ref{sec:meth}, whereas it's applicability is presented in Section~\ref{sec:exp}. Related work is discussed on Section~\ref{sec:rw}. Finally, we conclude the paper in Section~\ref{sec:concl}.

\section{Definitions and Background}
\label{sec:background}
In this section we present the definitions used throughout the paper (Section \ref{subsec:defn}). Next, we discuss existing models of popularity dynamics of individual objects (Section \ref{subsec:models}).

\subsection{Definitions} \label{subsec:defn}

We define an {\bf object} as a piece of media content stored on an
application.  Specifically,  an object on YouTube is a video, whereas, on an online radio like LastFM, we consider  (the webpage of) an artist as an object.  We also define an object on Twitter as a hashtag or a {\em musictag}\footnote{Users informing their followers which artists they are listening to.}.
A {\bf social activity} is the act of accessing -  posting, re-posting, viewing or listening to - an object on a social media application.  The {\bf popularity} of an object is the aggregate behavior of social activities on that  object. We here study popularity in terms of the most general  activities in each application: number of views  for YouTube videos, number of  plays for  LastFM artists,  and number of tweets with a hashtag. The popularity of an object is the sum of  \textbf{audience} (user's first visit) and, \textbf{revisits}, (or returning users), and the evolution of the popularity of an object over time defines a {\bf time series}. 

\subsection{Existing models of object popularity dynamics} \label{subsec:models}

{\bf Epidemic models:} Previous work on information propagation on online social networks (OSNs)  has exploited epidemic models~\cite{Hethcote2000} to explain the dynamics of the propagation process. An epidemic model describes the transmission of a  ``disease"  through individuals. The simplest epidemic model is the Susceptible-Infected (SI) model.
The SI model considers a fixed population divided into $S$ susceptible individuals and $I$ infected individuals.  Starting with $S(0)$ susceptible individuals  and $I(0)$ infected individuals, at each time step $\beta S(t-1)I(t-1)$ individuals get infected, and transition from the  $S$ state to the $I$ state. The product $S(t-1)I(t-1)$ accounts for all the possible connections between individuals. The parameter $\beta$ is the strength of the infectivity, or virus. 

%the evolution of the model is governed by the dynamics of at each time step, $\beta S(t-1)I(t-1)$ individuals get infected, transitioning from state $S$ to state $I$. The product $SI$ accounts for all the possible connections between individuals. The parameter $\beta$ is the strength of the infectivity, or virus.
Cha {\it et. al} used an SI model to study how information (i.e., the ``disease'') disseminates through social links on Flickr~\cite{Cha2008}, whereas Matsubara {\it et. al}~\cite{Matsubara2012} proposed an alternative model called SpikeM. SpikeM builds on an SI model by adding, among other things, a decaying power law infectivity per newly infected individual, which produces a behavior that is similar to the model proposed in~\cite{Crane2008}. The SpikeM model was used to captured the time series popularity for a single cascade. 
One of the reasons why the SI model is useful  to represent  online cascades of information propagation is that individuals usually  do not delete their posts, tweets or favorite markings~\cite{Cha2008,Matsubara2012}. Thus, once an individual is infected he/she remains infected forever (as captured by the SI model). 
%However, when considering popularity in general the ``forever infected'' may be false~\cite{Ribeiro2014,Anderson}. 

\begin{table}[ttt]
\scriptsize
\centering
\caption{Comparison of \model with other approaches}
\begin{tabular}{@{}lcccc@{}} 
\toprule
& Revisits & Non-Linear & Forecasting & Multi Cascade \\
\midrule
SI~\cite{Hethcote2000}         & & \checkmark   & & \\
SpikeM~\cite{Matsubara2012}     & & \checkmark & \checkmark & \\
TemporalDynamics~\cite{Radinsky2012} & & & \checkmark & \\
\midrule
\model  & \phantom{12345678} \checkmark \phantom{12345678} & \phantom{12345678} \checkmark \phantom{12345678} & \phantom{12345678} \checkmark \phantom{12345678} & \phantom{12345678} \checkmark \phantom{12345678} \\
\bottomrule
\label{tab:salesmat}
\end{tabular}
\vspace{-.8cm}
\end{table}

\noindent {\bf Temporal Dynamics Models:} Other models that can be explored in the study of content popularity dynamics are auto-regressive models and state space models, such as the Holt-Winters model and its extensions~\cite{Radinsky2012}. However, these models are linear in nature, and thus cannot account for more complex temporal dynamics observed in online content~\cite{Matsubara2012}. Although, these models have been successful in predicting {\it normalized} query behavior in search engines~\cite{Radinsky2012}, the descriptive power of such models is less attractive. For example, Holt-Winters based models are very general, that is, they are used to predict time series behavior, but will not take into account cascades, revisits or information dissemination. From a descriptive point of view, these models are of little help to understand the actual process that drives popularity evolution.

\noindent {\bf Multiple Cascades: } Very recently, the work of Hu {\it et. al} focused on the defining longevity of social impulses, or multiple cascades~\cite{Hu2014}. However, unlike our approach, the authors are not focused on modeling the long term popularity of objects. 

%We here capture multiple cascades in the \model model. To do this, we make use of a peak detection algorithm~\cite{Du2006}, which we adapted to estimate the the peak of the cascade and when it began. The \model model works well with this approach, and we leave the use of very recently proposed alternatives~\cite{Hu2014} as future work. 

Table~\ref{tab:salesmat}  summarizes the key properties of the aforementioned models as well as of our new \model model. In comparison these approaches, \model explicitly captures both revisits and multiple cascades, allows for non-linear solutions, and can be used for accurate forecasting. The next section presents the effect of revisits in both long and short term content popularity evolution for real world datasets. This is followed by the definition of the \model model.

%Note that no existing model explicitly captures the revisits by the same user of the same object. Yet,  as we show in  the next section,  the analysis of both long and short term content popularity evolution  in our datasets reveals that users often revisit the same object many times. In comparison these approaches, \model explicitly captures both revisits and multiple cascades, allows for non-linear solutions, and can be used for accurate forecasting.

%\section{Datasets}
%\label{sec:data}
%\input{030data}

\section{Content Revisit Behavior in Social Media}
\label{sec:charact}
We now analyze the revisit behavior in various social media applications. We first describe the datasets used in this analysis as well as in the evaluation of our model, and then discuss our main characterization findings.

\subsection{Datasets}

Our study is performed on  four large social activity datasets:
\begin{itemize}
    \item The Million Musical Tweets Dataset (MMTweet): consists of 1,086,808 tweets of users about artists they are listening to at the time~\cite{Hauger2013}. We focus on the artist of each tweet as an object. 25,060 artists were mentioned in tweets.
    \item The 2010 LastFM listening habits dataset (LastFM): consists of the whole listening habits (until May 5th 2009) of nearly 1,000 users, with over 19 million activities on  107,428 objects (artists)~\cite{Celma2010}.
    \item The 476 million Twitter tweets corpus (Twitter): accounts for roughly 20\% to 30\% of the tweets from June 1 2009 to December 31 2009~\cite{Yang2011}, and includes over 50 million objects (hashtags) tweeted by 17 million users.
    \item The YouTube dataset: Recently, YouTube began to provide the full daily time series (known as insight data) of visits for videos in the page of each video. We crawled the time series of over 3 million YouTube videos similar to as done in~\cite{Figueiredo2011}.
\end{itemize}

\subsection{Main findings}

 Our goal is to analyze how the popularity acquired by different objects, in the long and short runs,  is divided into audience and revisits.  In particular, we aim at assessing to which extent the number of revisits may be {\it larger} than the size of the audience, in which case popularity is largely a sum of repeated user activities.  Since this property may vary depending on  the type of content, we perform 
our characterization on the LastFM, MMTweet, and Twitter datasets. We leave the YouTube dataset out of this analysis since, unlike the other datasets, it does not contain individual social activities, but only popularity time series. We will make use of the YouTube dataset to fit and evaluate our \model model, in the next section.

\begin{figure}[ttt]
\parbox{.45\columnwidth}{
    \centering
    \includegraphics[scale=1]{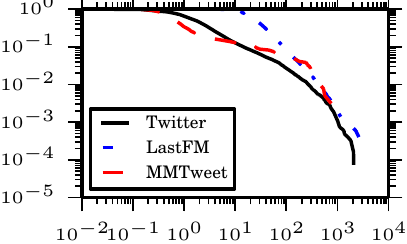}
    \caption{Distributions of $\frac{\#Revisits}{Audience}$.} 
    \label{fig:rev-over-audi}
}
\hspace{.3cm}
\parbox{.48\columnwidth}{
\centering
\scriptsize
\captionof{table}{Relationships between revisits, audience and  popularity.}
\begin{tabular}{llccc}
\toprule
 Dataset && Median & Median &  \% objects with \\
 &&  $\frac{\#Revisits}{Audience}$ &  $\frac{\#Revisits}{Popularity}$ & $\frac{\#Revisits}{Audience} > 1$ \\
\midrule
Twitter        && $1.70$  & $0.62$ & $66\%$  \\
MMTweet        && $0.68$  & $0.40$ & $33\%$  \\
LastFM         && $25.39$ & $0.96$ & $100\%$ \\
\bottomrule
\label{tab:audi-rev}
\end{tabular}
}
\vspace{-.5cm}
\end{figure}

We first analyze the distribution of the final values\footnote{Values computed at the time the data was collected.}  of popularity, audience, and revisits across objects in each dataset.  For illustration purposes, Figure~\ref{fig:rev-over-audi} shows the complementary cumulative distribution function of the ratio of the number of revisits to the audience size for all datasets, computed for objects with popularity greater than $500$. We filtered out very unpopular objects, which attract very little attention during the periods of our datasets (over 6 months each). Note that the probability of an object having more revisits than audience (ratio greater than 1) is  large. Indeed, though rare, the ratio of revisits to audience size reaches $10^{2}$ and even $10^{3}$. 

In order to better understand these findings across all datasets,  Table~\ref{tab:audi-rev} shows, for each dataset: (1) the median of the ratio of number of revisits to audience size, (2)  the median of the ratio of number of revisits to total popularity; and (3)  the percentage of objects where the revisits dominate the popularity  (i.e., ratio of number of revisits to the audience size  greater than 1).   Note that revisits dominate popularity in 66\% of the Twitter objects. Moreover, on median, 62\% of the total popularity of these objects is composed of revisits, which account  for 1.7 times more activities than the visits by new users (audience size).  Again,  for LastFM artists, revisits are over 25 times more frequent than the  visits by new users (on median), and the revisits dominate popularity in {\it all} objects.  In contrast, the ratios of number of revisits to audience size and to total popularity are smaller for MMTweet objects, most likely because users do not tweet about artists they are listening to all the time, but rather only when they wish to share this activity with their followers.  Yet, the revisits dominate popularity in 33\% of the objects.  These results provide evidence that, at least in the long run,  revisits are much more common than new users  for many objects in different applications.   For microblogs, though less intense, this behavior is  still non-negligible.  

\begin{table}[t]
\scriptsize
\centering
\caption{Quartiles of  the ratio $\frac{\#Revisits}{Audience}$ for various time windows $w$.}
\begin{tabular}{llccc} 
\toprule
Dataset & Time window ($w$) & \phantom{abc} $25^{th}$ percentile \phantom{abc} & \phantom{abc} Median \phantom{abc} & \phantom{abc} $75^{th}$ percentile \phantom{abc}  \\
\midrule
\multirow{4}{*}{Twitter}                 & 1 hour    & $1.08$ & $3.93$ & $12$   \\
                                         & 1 day     & $1$    & $2.5$  & $6.28$ \\
                                         & 1 week    & $0.66$ & $1.69$ & $4.28$ \\
                                         & 1 month   & $0.56$ & $1.44$ & $3.75$ \\
\midrule
\multirow{4}{*}{MMTweet}                 & 1 hour    & $0.25$ & $0.66$  & $12.5$  \\
                                         & 1 day     & $0.55$ & $0.83$  & $1.26$  \\
                                         & 1 week    & $0.41$ & $0.73$  & $1.41$  \\
                                         & 1 month   & $0.31$ & $0.56$  & $1.17$ \\
\midrule
\multirow{4}{*}{LastFM}                  & 1 hour    & $20$   & $21$   & $25$\\
                                         & 1 day     & $21$   & $28$   & $41$\\
                                         & 1 week    & $20$   & $30.5$ & $55.25$ \\
                                         & 1 month   & $14$   & $25$   & $48$ \\
\bottomrule
\label{tab:audi-rev-twindows}
\end{tabular}
\vspace{-.8cm}
\end{table}

We further analyze the effect of revisits on popularity, focusing now in the short term, by zooming  into smaller time windows $w$. 
Specifically, we analyze the distributions  of the ratios of number of revisits to audience size computed for window sizes $w$ equal to 
one hour, one day, one week, and one month. Table~\ref{tab:audi-rev-twindows} shows the three distribution quartiles for the various window sizes and datasets considered. These quartiles were computed considering only window sizes during which the popularity acquired by the object exceeds 20. We adopted this threshold to avoid biases in time windows with very low popularity, focusing on the periods where the objects had a minimal attention (note that 20 is still small considering that each trace has millions of activities).  

Focusing first on the LastFM dataset, we note that, regardless of the time window size, the number of revisits is at least one order of magnitude  (14x)  larger than the audience size for at least 75\% of the analyzed windows ($25^{th}$ percentile). In fact, the ratio between the two measures exceeds 55 for 25\% of the windows ($75^{th}$ percentile) on the weekly case. In contrast, in the MMTweet dataset, once again, the ratios are  much smaller. Nevertheless, at least  25\% of the of the windows we observe a burst of revisits in very short time, with the ratio exceeding 12 for the hourly cases.  Once again, we suspect that these lower ratios may simply reflect that users do not tweet about every artist they list to. 
Thus, in general, we have strong evidence that, for music-related content, popularity is mostly governed by revisits, as opposed to new users (audience).  

The same is observed, though with less intensity, in the Twitter dataset.  Revisits are more common than new users in 50\% of the time windows, for all sizes considered.  Indeed, considering hourly time windows, popularity is dominated by revisits for 75\% of the cases. While large ratios, such as those observed  for LastFM,  do not occur,  the number of revisits  can still be  12 times larger than the audience size during a single hour in 25\% of the Twitter hourly windows. 

{\bf Summary of findings: } Our main conclusions so far are: (1) for most objects in the Twitter and LastFM datasets, popularity, measured  both in the short (as short as 1 hour periods) and long runs, is mostly due to revisits than to audience size; and (2) revisits are less common on the MMTweet dataset, which we believe is due to data sparsity, but are still a significant component of the popularity acquired by a large fraction of the objects (in both long and short runs). These findings motivate the need for models that explicitly account for revisits in the popularity dynamics, which we discuss next.

\section{The \model Model}
\label{sec:meth}
In this section we introduce the proposed \model model(Section~\ref{sec:pr-model}), show how we fit the model to a given popularity time series (Section~\ref{sec:pr-fit}). In the next section we present results on the efficacy of the model on our datasets when compared to state-of-the-art alternatives, and the applicability of the  \model model. 

{\bf Notation: } We present vectors ($\mathbf{x}$) in bold. Sets are shown in non-bold calligraphy letters ($\mathcal{X}$), and variables are represented by lower case letters or Greek symbols ($x, \beta$). Moreover, $\mathbf{x}(i)$ means data index $i$ (from 1), and $\mathbf{x}(:i)$ means sub-vector up to $i$.

\subsection{Deriving the Model} \label{sec:pr-model}

The \model model is built based on the `Susceptible-Infected-Recovery' (SIR) compartments, extending for revisits and multiple cascades. Specifically, it captures the following behavior {\it for each individual object}:
\begin{itemize}
\item We assume a fixed population of individuals, where each individual can be in one of three states: susceptible, infected and recovered.
\item At any given time $s_{i}$, an external shock $i$ causes initial interest in the object. The shock can be any event that draws attention to the object, such as a video being uploaded to YouTube,  a news event about a certain subject, or even a search engine indexing a certain subject for the same time (thus making an object easier to be found). We assume that the initial shock  $s_1$ is always caused by one individual.  
\item New individuals discover the object by being infected by the first one. Moreover, after discovery, these ``newly infected'' can also infect other individuals, thus contributing to the propagation. 
\item Infected individuals may access (watch, play or tweet) the object. It is important to note that being infected does not necessarily imply in an access. For example, people may talk about a trending video before actually watching it. Each infected individual  accesses the object following a Poisson process  with rate $\omega$ ($\omega > 0$)\footnote{Both~\cite{Anderson,Huang2007} show the poissonian behavior of mutiple visits from the same user.}.
\item After some time, individuals lose interest in the object, which, in the model, is captured by a recovery rate $\gamma$. 
\item Multiple external shocks may occur for a single object. 
%Shocks are identified from the time series. We assume that the the population of individuals affected by different shocks do not interact (see below).
\end{itemize}

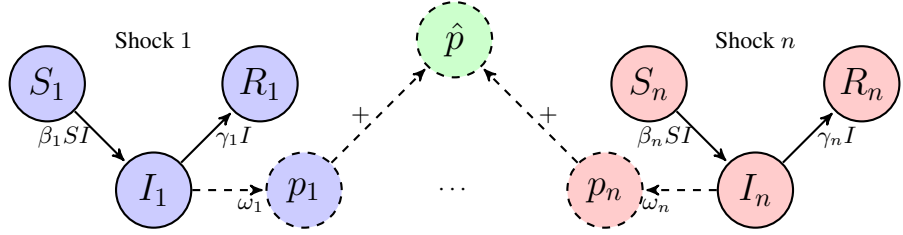
\begin{figure}[t]
\begin{tikzpicture}[%
    ->,>=stealth',shorten >=1pt,auto,node distance=2cm,thick,%
    main node/.style={%
        minimum size=1cm,%
        inner sep=0pt,%
        circle,%
        fill=blue!20,%
        draw,%
        font=\sffamily\Large\bfseries%
    }]

  \node[main node] (I1) {$I_1$};
  \node[main node] (S1) [above left of=I1] {$S_1$};
  \node[main node] (R1) [above right of=I1] {$R_1$};
  \node[main node, dashed] (p1) [right of=I1] {$p_1$};
  \node (l1) [above of=I1] {Shock $1$};

  \node (b1) [right of=I1] {};
  \node (b2) [right of=b1] {$\cdots$};
  \node (b3) [right of=b2] {};
  
  %\node (dots) [below of=b2] {$\cdots$};
  
  \node[main node, fill=red!20] (I2) [right of=b3] {$I_n$};
  \node[main node, fill=red!20] (S2) [above left of=I2] {$S_n$};
  \node[main node, fill=red!20] (R2) [above right of=I2] {$R_n$};
  \node[main node, dashed, fill=red!20] (p2) [left of=I2] {$p_n$};
  \node (l2) [above of=I2] {Shock $n$};
  
  \node[main node, dashed, fill=green!20] (ph) [above of=b2] {$\hat{p}$};

  \path[every node/.style={font=\sffamily\small}]
    (S1) edge node [left]  {$\beta_1 SI$} (I1)
    (I1) edge node [right] {$\gamma_1 I$} (R1)
    (S2) edge node [left]  {$\beta_n SI$} (I2)
    (I2) edge node [right] {$\gamma_n I$} (R2)
    (I1) edge [dashed] node [below right] {$\omega_1$} (p1)
    (I2) edge [dashed] node [below left]  {$\omega_n$} (p2)
    (p1) edge [dashed] node [left] {$+$} (ph)
    (p2) edge [dashed] node [right] {$+$} (ph);
\end{tikzpicture}
\caption{Individual shocks that when added up account for the \model model.} 
\label{fig:model}
\vspace{-.5cm}
\end{figure}

Figure~\ref{fig:model} presents the \model model. In the figure, three compartments are shown for each shock $S_i$, $I_i$, and $R_i$, which represent the number of susceptible, infected and recovered individuals for shock $i$, respectively. Variable $p_i$, associated with shock $i$,  measures the popularity acquired by the object due this shock.  The total popularity of the object, i.e., the sum of the values of $p_i$ for all shocks, is denoted by $\hat p$.  We first present the model for a single shock, and then generalize the solution for multiple shocks. Also, we drop the subscripts while discussing a single shock. We present the model assuming discrete time, referring to each time tick as a time window.

Each shock begins with a given susceptible population ($S(0)$) and one infected individual ($I(0) = 1$). The total population is fixed and given by ($N = S(0) + 1$). The $R$ compartment captures the individuals that lost interest in the object. Similarly the SI model, $\beta SI$ susceptible individuals become infected in each time window. Moreover,  $\gamma I$ individuals loose interest in (i.e., recover from)  the object in each window. Revisits  to the object  are captured by   the rate $\omega$. Thus $\omega$ is  the expected number of accesses of an individual up to time $t$, the probability of the individual accessing the object $k$ times during a time interval  of $\tau$ windows is given by $ P(v(t + \tau) - v(t) = k) = \frac{(\omega \tau)^k e^{-\omega \tau}}{k!}$. 

We assume that the shock starts at time zero, thus focusing the dynamics {\it after} the shock. Under this assumption, the equations that govern a single shock are:
\begin{align}
    S(t) &= S(t - 1) -\beta S(t - 1) I(t - 1) \\
    I(t) &= I(t - 1) + \beta S(t - 1) I(t - 1) - \gamma I(t - 1) \\
    R(t) &= R(t - 1) + \gamma I(t - 1) \\
    p(t) &= \omega  I(t).
\end{align}

The equation  $p(t) = \omega I(t)$ accounts for the expected number of  times infected individuals  access the object, thus capturing the popularity of the object at time $t$ due to the shock. We can also define the expected audience size of  the object at time $t$ due to the shock, $a(t)$,  as: $a(t) = (1 - e^{-\frac{\omega}{\gamma}}) \beta S(t - 1) I(t - 1)$. Each newly infected individual ($\beta S(t - 1) I(t - 1)$) will stay infected for $\gamma^{-1}$ windows (see~\cite{Hethcote2000}). The probability of generating at least one access while the individual is infected is: $1 - P(v(t + \gamma^{-1}) - v(t) = 0) = 1 - e^{-\frac{\omega}{\gamma}}$. Thus, we here capture the individuals which where infected at some time and generated at least one access.

The \model model is thus defined as the sum of the popularity values due to multiple  shocks.  We discuss how to determine the number of shocks in the next section. Given a set of shocks $\mathcal{S}$, where shock $i$  starts at   given time $s_i$, the popularity $\hat{p}$ is: 
\begin{align}
    \label{eq:model}
    \hat p(t) = \sum_{i, s_{i} \in \mathcal{S}} p_i(t - s_i) \mathbbm{1}[t > s_{i}]
\end{align}
\noindent where $\mathbbm{1}[t > s_{i}]$ is an indicator function that takes value of 1 when $t > s_{i}$, and 0 otherwise.  Audience, size $\hat{a}(t)$ can be similarly defined. Also, both in the single shock and in the  \model models, the number of revisits  at time $t$, $\hat r(t)$, can be computed as $\hat r(t) = \hat p(t) - \hat a(t)$. The overall population that can be infected is defined by $N = \sum_i N_i = \sum_i S(0)_i + 1$.

Note that we assume that the population of different shocks do not interact, that is, an infected individual from shock $s_i$ does not interact with a susceptible one from shock $s_j$, where $i \neq j$. While this may not hold  for some objects (e.g.,  people may  hear about the same content from two different populations), it may be a good approximation for objects that become popular in  large scale (e.g., objects that are propagated world wide). It also allows us to have different $\beta_i$, $\gamma_i$, and $\omega_i$ values for each population. Intuitively, the use of different parameters for each shock captures the notion that some objects may be more (or less) interesting for different populations. For example, samba songs may attract more interest from  people  in Brazil. 

{\bf Adding a period:} In some cases, the popularity of an object may be affected by periodical factors. For example, songs may get more plays on weekends. We add a period to the \model model by making $\omega$ fluctuate in a periodic manner. That is:
\begin{align}
    \omega_i(t) = \omega_i * (1 - \frac{m}{2} * (sin(\frac{2 \pi (t + h)}{e}) + 1)).
\end{align}
\noindent $e$ is the period, and $sin$ is a sine function. For example, for daily series we may set $e=7$ if more interest is expected on weekends. Since an object may have been uploaded on a Wednesday, we   use the shift $h$ parameter to correct the sine wave to peak on weekends. The amplitude $m$ captures oscillation in visits. The same period parameters are applied to every shock model.  This approach is similar to the one adopted in \cite{Matsubara2012}.

The final \model model will have 5 parameters to be estimated from the data {\it for each} shock, namely, $S(0)_i$, $\beta_i$, $\gamma_i$, $\omega_i$, $s_i$; plus the $m$  and $h$ period parameters. The last two do not change for individual shocks. We decided to fix $e$ in our experiments to $7$ days, when using daily time windows, and $e=24$ hours when using hourly series.

\subsection{Fitting the Model} \label{sec:pr-fit}

\begin{algorithm}[t!]
\caption{Fitting the \model model.  Only the time series is required as input.}% The algorithm needs only a time series $\mathbf{t}$ as input. The $LM$ provides the least squares fit; $PhoenixR$ evaluates the model; and, $Cost$ computes the MDL cost.}
\label{algo:fit}
\scriptsize
\begin{algorithmic}[1]
\Function{FitPhoenixR}{$\mathbf{t}$}
 
    \State{$\epsilon = 0.05$}
    \State{$\mathbf{s} \gets \{\}$}
    \State{$\mathbf{p}, \mathbf{s'} \gets FindPeaks(\mathbf{t})$} 
    
    \State{$\mathbf{s}[1] = 0$}
    \State{$\mathbf{s} \gets append(\mathbf{s'})$}
    \State{$\mathcal{P} \gets \{\}$}
    \State{$min\_cost \gets \infty$}   
    \For{$i \gets 1 \quad \text{to} \quad |s|$}
        
        \State{$\mathcal{F} \gets LM(\mathbf{t}, \mathbf{s}(:i))$}
        \State{$\mathbf{m} \gets PhoenixR(\mathcal{F})$}
        \State{$mdl\_cost \gets  Cost(\mathbf{m}, \mathbf{t}, \mathcal{F})$}

        \If{$mdl\_cost < min\_cost$}
            \State{$min\_cost \gets mdl\_cost$}
            \State{$\mathcal{P} \gets \mathcal{F}$}
        \EndIf

        \If{$mdl\_cost > min\_cost * (1 +\epsilon)$} 
            \State{\textbf{break}}
        \EndIf
    \EndFor
    \State \Return{$\mathcal{P}$}
\EndFunction
\end{algorithmic}
\end{algorithm}

We now discuss how to fit the \model parameters to real world data. Our goal is to produce a model that delivers a good trade-off between parsimony (i.e., small number of parameters) and accuracy.  To that end, three issues must be addressed: (1) the identification of the start time  of each individual shock; (2) an estimation of the cost of  the model associated with multiple shocks; and, (3) the fitting algorithm itself.  Note that  one key component of the fitting algorithm is model selection: it is responsible for determining the number of shocks that will compose the \model model, choosing a value based on the  cost estimate and model accuracy. 

\textbf{Finding the start times $s_i$ of the shocks:} Intuitively, we expect each shock to correspond to a peak in the time series. Indeed, previous work has looked at the dynamics of  single shock cascades, finding a single prominent peak in each cascade~\cite{Matsubara2012,Bauckhage2013}. With this in mind, instead of searching for $s_i$ directly, we initially attempt to find peaks. We can achieve both tasks using a continuous wavelet transform based peak finding algorithm~\cite{Du2006}.  We chose this algorithm since it has  the following key desirable   properties. Firstly, it can find peaks regardless of the ``volume'' (or popularity in the present context) in the time windows surrounding the peaks. It does so by only considering peaks with a high signal to noise ratio in the series, that is, peaks that can be distinguished the time series signal around the candidate peak.  Secondly, the algorithm is fast, with complexity in the order of the length, $n$, of the time series ($O(n)$). Lastly and more importantly, using the algorithm we can estimate both the peaks and the start times of the shocks that caused each peak. We shall refer to the algorithm as $FindPeaks$.

As stated  $FindPeaks$  makes use of a continuous wavelet transform to find the peaks of the time series.  Specifically, we apply the Mexican Hat Wavelet\footnote{\url{https://en.wikipedia.org/wiki/Mexican_hat_wavelet}} for this task. The Mexican Hat Wavelet is parametrized by a half-width $l$. We use half-widths ($l$) of values $\{1, 2, 4, 8, 16, 32, 64, 128, 256\}$ to find the peaks. Thus, for the peak identified at position $k_i$, with wavelet determined by the parameter $l_i$, we define  the start point of the shock $s_i$ as: $s_i = k_i - l_i$. We found that using the algorithm with the default parameters presented in~\cite{Du2006}, combined with our MDL fitting approach (see below), proved accurate in modeling the popularity of objects\footnote{We used the open source implementation available with SciPy (\url{http://scipy.org})}.

\textbf{Estimating the cost of the model with multiple shocks:}  we estimate the cost of a model with $|\mathcal{S}|$ shocks  based on the minimum description length (MDL) principle \cite{Nannen2010,Hansen2001}, which is largely  used for problems of model selection. To apply the MDL principle, we need a coding scheme that can be used to compress both the model parameters and the likelihood of the data given the model.
We here provide a new intuitive coding scheme, based on the MDL principle, for describing the \model model with $|\mathcal{S}|$  shocks, assuming a popularity time series of $n$ elements (time windows). As a general approach, we code natural numbers using the $\log^*$ function (universal code length for integers)\footnote{$\log^*(x) = 1 + \log^*(\log x)$ if $x > 1$. $\log^*(x) = 1$ otherwise. We use base-2 logarithms.}~\cite{Hansen2001}, and fix the cost of floating point numbers at $c_f = 64$ bits.

For each shock $i$, the  complexity of the description of the  set of parameters  associated with $i$  consists of the following terms: $\log^*(n)$ for the $s_i$ parameter (since the start time of $i$ can be at any point in the time series); $\log^*(S_i(0))$ for the initial susceptible population; and $3 * c_f$ for $\beta_i$, $\gamma_i$, and $\omega_i$. We note that an additional cost of $\log^*(7) + 2 * c_f$ is incurred if a period is added to the model. However, we ignore this component here since it is fixed for all models.  Therefore, it does not affect model selection. The cost associated with the set of parameters $\mathcal{P}$ of   all   $|\mathcal{S}|$ shocks is: 
\begin{align}
    Cost(\mathcal{P}) = |\mathcal{S}| \times  (\log^*(n) + \log^*(S_i(0)) + 3 * c_f) + \log^*{|\mathcal{S}|}.
\end{align}

Given the full parameter set $\mathcal{P}$, we can encode the data  using Huffman coding, i.e., a number of bits is assigned to each value which is the logarithm of the inverse of the probability of the values  (here, we use a Gaussian distribution as suggested in~\cite{Nannen2010} for the cases when not using probabilistic models.). 

Thus, the cost associated with coding of the time series given the parameters is:
\begin{align}
    Cost(\mathbf{t}\mid\mathcal{P}) = -\sum_{i=1}^n \log(p_{gaussian}(\mathbf{t}(i) - \mathbf{m}(i); \mu, \sigma)).
\end{align}
\noindent  where $\mathbf{t}$ is the time series data and  $\mathbf{m}$ is the  time series  produced by the model (i.e., $\mathbf{t}(i) - \mathbf{m}(i)$ is the error of the model at time window $i$.) Here, $p_{gaussian}$ is the probability density function of a Gaussian distribution with mean $\mu$ and standard deviation $\sigma$  estimated from the model errors. We do not include the costs of encoding $\mu$ and $\sigma$ because, once again, they are constant for all models. The total cost is:
\begin{align}
Cost(\mathbf{t}; \mathcal{P}) = \log^*{n} + Cost(\mathcal{P}) + Cost(\mathbf{t}\mid\mathcal{P}).
\end{align}
\noindent This accounts for the parameters cost, the likelihood cost, and the cost of the data size.

\textbf{Fitting algorithm:}  The model fitting approach  is summarized in Algorithm~\ref{algo:fit}. The algorithm receives as input a popularity time series $\mathbf{t}$. It first identifies candidate shocks using the $FindPeaks$ method, which returns the peaks $\mathbf{p}$ and  the start times  $\mathbf{s'}$ of the corresponding shocks  in decreasing order of  peak volume  (line 3).
To account for the upload of the object, we include one other shock starting at time $s_1 = 0$, in case a shock was not identified in this position. Each $s_i$ is stored in vector $\mathbf{s}$,  ordered by the volume of the each identified peak (with the exception of $s_1 = 0$ which is always in the first position) (lines 4 and 5).
We then fit the \model model using the Levenberg-Marquardt  (LM) algorithm adding one shock at a time, in the order they appear in $\mathbf{s}$ (loop in line 9), that is, in decreasing order of peak volume (after the initial shock). 
Intuitively, shocks that lead to larger peaks account for more variance in the data. For each new shock added, we evaluate the MDL cost (line 12).
We keep adding new shocks as long as the MDL cost decreases (line 13) or provided that an increase of at most $\epsilon$ over the best model is observed\footnote{MDL based costs will decrease with some variance and then increase again. The $\epsilon$ threshold is a guard against local minima due to small fluctuations.} (line 17). We set the Levenberg-Marquardt algorithm to evaluate the mean squared errors of the model and adopt a threshold $\epsilon$ equal to 5\%.
We also note that we initialize each parameter randomly (uniform from 0 to 1), except for $S_i(0)$ values. For the first shock we do test multiple initial values: $S_1(0)=$ $10^3$, $10^4$, $10^5$, and $10^6$. The other $S_i(0)$ values are initialized to the corresponding peak volume.

\section{Experiments}
\label{sec:exp}
In this section we discuss the experimental evaluation of the \model model. Initially, we present results on the efficacy of the model on our datasets when compared to state-of-the-art alternatives (Section~\ref{sec:pr-compare}) Next, we show results for the applicability of the model for popularity prediction (Section~\ref{sec:app})\footnote{All of our source code is provided at: \url{http://github.com/flaviovdf/phoenix}}. 

\subsection{Is \model Better than Alternatives?} \label{sec:pr-compare}

We compare \model with two state-of-the-art alternatives: the TemporalDynamics~\cite{Radinsky2012}, used to model query popularity; and the SpikeM model~\cite{Matsubara2012}, which captures single cascades.  We compare these models in terms of time complexity, accuracy, estimated by the root mean squared errors (RMSE),  and  cost-benefit. For the latter, we use the Bayesian Information Criterion (BIC)~\cite{Radinsky2012},  which captures the tradeoff between cost (number of parameters) and accuracy of the model.

\begin{figure}[t]
    \centering
    \includegraphics[scale=.6]{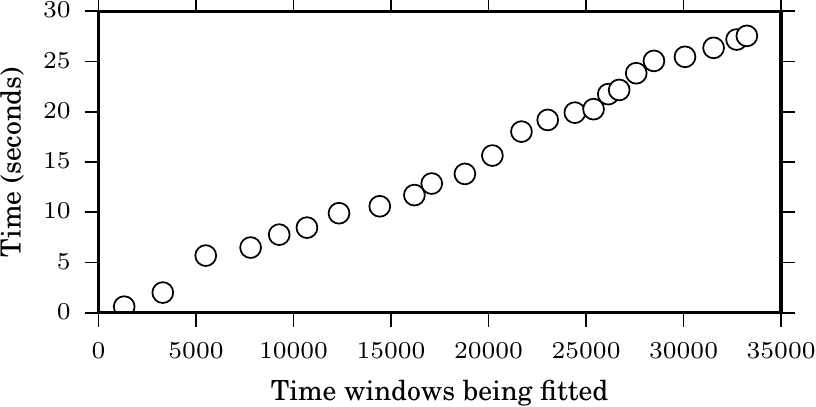}
    \caption{Scalability of \model}
    \label{fig:scale}
    \vspace{-.3cm}
\end{figure}

In terms of time complexity, we note that the \model model scales linearly with the length of the time series $n$. This is shown in Figure~\ref{fig:scale}, which presents the number of seconds ($y$-axis) required to fit a time series with a given number of time windows ($x$-axis). TemporalDynamics also has linear time complexity \cite{Radinsky2012}. In contrast, the equations that govern the SpikeM model requires quadratic ($O(n^2$)) runtime on the time series length, making it much less scalable to large datasets.   

In terms of accuracy, we make an effort to compare \model with the alternatives in fair settings, with datasets with similar characteristics from those used in the original papers. In particular, when comparing with TemporalDynamics, we run the models proposed in \cite{Radinsky2012} selecting the best one (i.e., the one with smallest root mean squared error) for each time series. Moreover, we use long term daily time series (over 30 days), with a total popularity of at least 1,000\footnote{ Similar results were achieved using other thresholds.}. We compare \model and TemporalDynamics under these settings in our four datasets, including YouTube.

When comparing with SpikeM, we use Twitter hourly time series trimmed to 128  time windows around the largest peak (most popular hour). We focus on the 500 most popular of these times series for comparison. We chose this approach since this is the same dataset explored by the authors. Moreover, we focus on a smaller time scale because the SpikeM was proposed for single cascades only. 

\begin{table}[t]
\scriptsize
\centering
\caption{Comparison of \model with TemporalDynamics~\cite{Radinsky2012} and SpikeM~\cite{Matsubara2012}: Average RMSE values (with 95\% confidence intervals in parentheses). Statistically significant results (including ties) are shown in bold.}
\begin{tabular}{lcccccc}
\toprule
                &&  \multicolumn{2}{c}{\model vs. TemporalDynamics (daily series)} &\phantom{abcdefg}& \multicolumn{2}{c}{\model vs. SpikeM (hourly series)} \\
                \cmidrule{3-7}
                &&   \phantom{a} RMSE \phantom{ab} & \phantom{a} RMSE \phantom{a} &&    \phantom{a} RMSE \phantom{a} & \phantom{a} RMSE \phantom{a} \\
                &   & \model \phantom{ab} & \phantom{ab} TemporalDynamics &&   \model \phantom{ab} & \phantom{ab} SpikeM \\
\midrule
MMTweet &&  {\bf 2.93}   ($\pm$ 0.23)   & 4.18    ($\pm$ 0.49)      &&   - & - \\
LastFM  &&   {\bf 7.09}   ($\pm$ 0.23)   & 8.31    ($\pm$ 0.32)     &&   - & - \\
Twitter &&   {\bf 72.05}  ($\pm$ 6.08)   & 194.79  ($\pm$ 20.49)    && {\bf 10.83} ($\pm$ 1.61) & {\bf 9.77} ($\pm$ 2.24)\\
YouTube &&   {\bf 280.58} ($\pm$ 29.29)  & 3429.19 ($\pm$ 577.76)   &&  - & - \\
\bottomrule
\label{tab:gains}
\end{tabular}
\vspace{-.8cm}
\end{table}

 Table~\ref{tab:gains} shows  the average RMSE (along with corresponding 95\% confidence intervals) computed over the considered time series for all models. Best results of each comparison (including statistical ties) are shown in bold.  Note that \model has statistically lower RMSE than TemporalDynamics in all datasets. These improvements come particularly from the non-linear nature of \model, which better fits the long term popularity dynamics of most objects. 
 The difference between the models is more striking for the YouTube dataset, where most time series cover long periods (over 4 years in some cases). The linear nature of TemporalDynamics largely affects its performance in those cases, as many objects do not experience a linear popularity evolution over such longer periods of time. 
As result,   \model produces reductions on average RMSE of over one order of magnitude.   In contrast, the 
 gap between both models is smaller in the LastFM dataset,  where the fraction of objects  (artists) for which a linear fit is reasonable is larger.
Yet, \model produces results that are still statistically better, with a reduction on  average RMSE of 15\%. 

When comparing with SpikeM, the \model model produces results that are statistically tied. We consider this result very positive, given that this comparison favors SpikeM: the time series cover only 128 hours, and thus there is no much room for improvements from capturing multiple cascades, one key feature of \model.  Yet, we note that our model is more general and suitable to modeling popularity dynamics in the longer run, besides being much more scalable, as discussed above.

As a final comparison, we evaluate the cost-benefit of the models using BIC, as suggested by \cite{Radinsky2012}. 
We found that we beat TemporalDynamics in terms of BIC on at least 80\% of the objects in all datasets but LastFM.
For LastFM objects, the reasonable linear evolution of popularity of many objects, makes the cost-benefit of TemporalDynamics superior. Yet, \model is still the preferred option in 30\% of the objects in this dataset.  Compared to SpikeM we also find that again, statistically equal BIC scores are achieved for both models.

\subsection{Predicting Popularity with \model} \label{sec:app}

\begin{table}[t]
\scriptsize
\centering
\caption{Comparing Phoenix-R with TemporalDynamics~\cite{Radinsky2012} for prediction. The values on the table are RMSE. Statistically significant results are in bold}
\begin{tabular}{@{}lcccccccccccc@{}}
\toprule
                %&&  \multicolumn{9}{c}{PhoenixR} & \phantom{abc} & \multicolumn{9}{c}{TemporalDynamics} \\
                %\cmidrule{3-21}
                & \phantom{abc} & \multicolumn{3}{c}{5\%} & \phantom{abc} & \multicolumn{3}{c}{25\%} & \phantom{abc} & \multicolumn{3}{c}{50\%} \\
\cmidrule{3-5} \cmidrule{7-9} \cmidrule{11-13}
                &\phantom{abc} & 1 & 7 & 30 & \phantom{abc} & 1 & 7 & 30 & \phantom{abc} & 1 & 7 & 30 \\
\midrule
\multirow{2}{*}{MMTweet}         &PhoenixR    & {\bf 11.61} & {\bf 12.78} & {\bf 15.15} && {\bf 8.67} & {\bf 6.74} & {\bf 8.82}   && {\bf 4.08} & {\bf 6.87} & {\bf 13.58} \\
                                 &TempDynamics& {\bf 17.07} & {\bf 17.41} & {\bf 16.52} && {\bf 9.63} & {\bf 10.78} & {\bf 14.46} && {\bf 25.19} & {\bf 23.08} & {\bf 30.39} \\
\midrule
\multirow{2}{*}{Twitter}         &PhoenixR    &  {\bf 53.68}  & {\bf 60.78}  & {\bf 215.76} && {\bf 132.21} & {\bf 135.15} & {\bf 210.30} && {\bf 75.58} & {\bf 229.59} & {\bf 254.93}\\
                                 &TempDynamics&  {\bf 104.45} & {\bf 129.36} & {\bf 255.69} && 643.39 & 643.83 & 786.50 && 420.74 & 587.86 & 598.75\\
\midrule
\multirow{2}{*}{LastFM}          &PhoenixR    & {\bf 2.37} & {\bf 3.97} & {\bf 5.71} && {\bf 8.60} & {\bf 12.06} & {\bf 14.66} && {\bf 11.34} & {\bf 15.03} & {\bf 15.43} \\
                                 &TempDynamics& 6.47 & 7.03 & 8.00 && 11.15 & 14.62 & 17.86 && 14.91 & 18.15 & 18.80  \\
\midrule
\multirow{2}{*}{YouTube}         &PhoenixR     & {\bf 91.62} & {\bf 106.38} & {\bf 138.88} && {\bf 83.76} & {\bf 113.14} & {\bf 147.04} && {\bf 127.53} & {\bf 97.97} & {\bf 115.97} \\
                                 &TempDynamics & 3560.65 & 3631.09 & 3661.81 && 5091.82 & 5107.82 & 5143.70 && 4136.14 & 4139.73 & 4169.26\\
\bottomrule
\label{tab:predictionres}
\end{tabular}
\vspace{-.8cm}
\end{table}

We here assess the efficacy of \model for predicting the popularity of objects a few time windows into the future, comparing it against  TemporalDynamics\footnote{We do not use SpikeM for this task, as it is suitable for tail forecasting only (i.e., predicting after the peak)}. To that end,   we train the \model and TemporalDynamics models for each time series using 5\%, 25\%, and 50\% of the initial daily time windows. We then use  the  $\delta$ time windows  following the training period as validation set  to learn  model parameters.  In each setting, we train 10 models  for each time series, selecting the best one on the validation period.  We then use the selected model to estimate the popularity of the object $\delta$ windows after the validation (test period).  We experiment with $\delta$ equal to   1, 7 and 30 windows.

Table~\ref{tab:predictionres} shows  the average RMSE of both models on the test period.  Confidence intervals are omitted for the sake of clarity, but the  best results (and statistical ties) in each setting 
are shown in bold. \model produces more accurate predictions than TemporalDynamics in practically all scenarios and datasets. Again, the improvements are quite striking for the YouTube dataset, mainly because the time series cover long periods (over 4 years in some cases).  While the linear TemporalDynamics model fits reasonably well  the popularity dynamics of some objects, it performs very poorly on others,  thus leading to high variability in the results. In contrast, \model is much more robust, producing more accurate  predictions for most objects, and thus being more suitable for modeling and predicting long periods of social activity.

\section{Related Work}
\label{sec:rw}
Popularity prediction of social media has gained a lot of attention recently, with many efforts focused on linear methods to achieve this task \cite{Pinto2013,Szabo2010,Radinsky2012}. However, not all of these methods are useful for modeling {\it individual} time series. For example, linear regression based methods \cite{Pinto2013,Szabo2010} can be used for prediction  but are not explanatory of individual time series. Moreover, as we showed in our experiments, there is strong evidence that linear methods are less suitable for modeling popularity dynamics than non-linear ones, particularly for long term dynamics. This comes from the non-linear behavior of social cascades~\cite{Matsubara2012}.  Li {\it et. al.}~\cite{Li2013} proposed a non-linear popularity prediction model. However they focused on modeling the video propagation through links on a single online social network, and not on general time series data, as we do here. 
%Moreover, unlike \model, their method does not account for revisits or multiple cascades.

Recent work has also focused on modeling the dynamics of news evolution~\cite{Matsubara2012}, or posts on news aggregators~\cite{Lerman2010,Lakkaraju2013,Bauckhage2013}. These prior efforts do not explicitly account for revisits nor multiple cascades, as we do. For example, the authors either assume unique visits only~\cite{Matsubara2012}, or focus on applications that do not allow revisits (e.g., once a user  likes a news posted on a application, she/he cannot like it a second time)~\cite{Lakkaraju2013}. In other cases, the models do not distinguish between a first visit by a user and a revisit~\cite{Bauckhage2013}.

Very recently, Anderson {\it et. al.}~\cite{Anderson} analyzed revisits in social media applications. However, unlike we do here, the authors were not focused on modeling the evolution of popularity of individual objects, but rather the aggregate and user behavior.

\section{Conclusions}
\label{sec:concl}
In this paper we presented the \model model for social media popularity time series. Before introducing the model, we showed the effect of revisits on the popularity of objects on large social activity datasets. Our main findings are:
\begin{itemize}
\item {\bf Discoveries: } We explicitly show the effect of revisits in social media popularity.
\item {\bf Explanatory model: } We define the \model, which explicitly accounts for revisits and multiple cascades. Factors not captured by state-of-the art alternatives.
\item {\bf Scalable and Parsimonious: } Our fitting approach make's use of the MDL principle to achieve a parsimonious description of the data. We also show that fitting the model is scalable (linear time). 
\item {\bf Effectiveness} of model: We showed the effectiveness of the model not only when describing popularity time series, but also when predicting future popularity values for individual objects. Gains can be up to one order of magnitude larger than baseline approaches, depending on the dataset.
\end{itemize}
As future work we intend on extending the \model model to deal with: (1) interacting populations between shocks; (2) multiple cascades from a single population; and, (3) fitting on multiple time series at once (e.g., audience and revisits). 
\blfootnote{{\bf Acknowledgments:} This research is in part funded by the Brazilian National Institute of Science and Technology for Web Research (MCT/CNPq/INCT Web Grant Number 573871/2008-6), and by the authors' individual grants from Google's Brazilian Focused Research Award, CNPq, CAPES and Fapemig. It was also supported by NSF grants CNS-1065133 and CNS-1314632, as well as ARL Cooperative Agreements W911NF-09-2-0053 and W911NF-11-C-0088. The views and conclusions contained in this document are those of the authors and should not be interpreted as representing the official policies, either expressed or implied of the NSF, ARL, or other funding parties. The U.S. Government is authorized to reproduce and distribute reprints for Government purposes notwithstanding any copyright notation here on.}

%The views and conclusions contained in this document are those of the author and should not be interpreted as representing the official policies, either expressed or implied of the NSF, ARL, or the U.S. Government. The U.S. Government is authorized to reproduce and distribute reprints for Government purposes notwithstanding any copyright notation hereon.
%   Any opinions, findings, and conclusions or recommendations expressed in this
%   material are those of the author(s) and do not necessarily reflect the views
%   of the National Science Foundation, Army Research Laboratory, or other funding parties.
%   The U.S. Government is authorized to reproduce and 
%   distribute reprints for Government purposes notwithstanding 
%   any copyright notation here on.
\bibliographystyle{abbrv}
\bibliography{BIB/myref}

\end{document}